\documentclass[aps,pra,preprint,floatfix]{revtex4-1}
\usepackage{graphicx,xcolor}
\usepackage{amsmath,amssymb}
\usepackage{dsfont}

\newcommand{\heff}{H_{\scriptscriptstyle \rm eff}}
\newcommand{\hs}{\mathcal{H}_{\scriptscriptstyle \rm S}}
\newcommand{\hdr}{\mathcal{H}_{\scriptscriptstyle \rm DR}}

\newcommand{\hl}{\mathcal{H}_{\scriptscriptstyle \rm L}}

\newcommand{\hsl}{\mathcal{H}_{\scriptscriptstyle \rm SL}}

\newcommand{\hdd}{\mathcal{H}_{\scriptscriptstyle \rm DD}}

\newcommand{\nn}{\nonumber}

\newcommand{\trl}{\text{Tr}_{ \scriptscriptstyle \rm L}}
\newcommand{\rh}{\rho_{\scriptscriptstyle  S}}
\newcommand{\rl}{\rho_{\scriptscriptstyle \rm L}^{\scriptscriptstyle \rm eq}}

\newcommand{\etal}{\textit{et al.\ }}
\newcommand{\sts}{\vert_{t\rightarrow\infty}}
\newcommand{\stoo}{\vert_{t\rightarrow 0}}

\newcommand{\comment}[1]{ }
\begin{document}

\title{Cascaded dynamics of a periodically driven dissipative dipolar system}

\author{Saptarshi Saha}
\email{ss17rs021@iiserkol.ac.in}

\author{Rangeet Bhattacharyya}
\email{ rangeet@iiserkol.ac.in}

\affiliation{ Department of Physical Sciences, 
Indian Institute of Science Education and Research Kolkata,
Mohanpur - 741 246, WB, India }
 
\date{\today}

\begin{abstract}

Recent experiments show that periodic drives on dipolar systems lead to long-lived prethermal states. These
systems are weakly coupled to the environment and reach prethermal states in a timescale much shorter than
the timescale for thermalization. Such nearly-closed systems have previously been analyzed using Floquet
formalism, which shows the emergence of a prethermal plateau. We use a fluctuation-regulated quantum master
equation (FRQME) to describe these systems. In addition to the system-environment coupling, FRQME
successfully captures the dissipative effect from the various local interactions in the system. Our
investigation reveals a cascaded journey of the system to a final steady state. The cascade involves a set of
prethermal or arrested states characterized by a set of quasi-conserved quantities. We show that these
prethermal states emerge in a timescale much shorter than the relaxation timescale. We also find and report
the existence of a critical limit beyond which the prethermal plateau ceases to exist.

\end{abstract}
\maketitle

Experimental evidence shows that nearly isolated systems exhibit prethermalization in a timescale much
shorter than that expected from the system-environment coupling \cite{berges_prethermalization_2004,
sen_quasi-equilibrium_2004}. A complete understanding of how such nearly-isolated systems reach a thermal
state remains an enigma. The understanding of the dynamics of this process is now at the forefront of
quantum, and statistical mechanics research \cite{deutsch_2018, ueda_2020}. Earlier, Deutsch showed an approach to
thermalization in a microcanonical ensemble might be realized through the inclusion of a random Hamiltonian in
the system \cite{deutsch_1991,deutsch_2018}. Another major milestone in the understanding of the thermalization
process came from Srednicki; he proposed separate thermalization within the subspace of each eigenvalue of
the system, known as the eigenstate thermalization hypothesis (ETH) \cite{srednicki_1994, srednicki_1999}. In
the process of thermalization, the system may encounter long-duration prethermal states at an intermediate
timescale. Such prethermal states are characterized by incomplete loss of initial memory, as opposed to the
complete loss in a thermalization process \cite{ueda_2020}. 

For isolated many-body quantum systems, under some circumstances, a periodic drive can lead the system to
the long-lived prethermal plateau. It has been theoretically analyzed using Floquet theory and is known as
Floquet prethermalization \cite{kuwahara_floquetmagnus_2016, abanin_exponentially_2015,
fleckenstein_prethermalization_2021, holthaus_floquet_2016, weidinger_floquet_2017, bukov_prethermal_2015,
yin_prethermal_2021}. The prethermal plateau has a large number of applications in quantum computation,
information processing, and quantum state engineering \cite{bukov_universal_2015}. Experimental
demonstrations of Floquet prethermalization are opening a new frontier in non-equilibrium quantum physics
\cite{rubio-abadal_floquet_2020, viebahn_suppressing_2021, beatrez_floquet_2021, peng_floquet_2021,
neyenhuis_observation_2017}.  

We note that a recently introduced fluctuation-regulated quantum master equation (FRQME) is capable of
describing the non-equilibrium dynamics of a nearly closed system in a unique way by taking into account
the dissipators from system interactions in addition to the system-environment coupling
\cite{chakrabarti2018b}. The unique feature in FRQME is the addition of an explicit bath fluctuation term in
the dynamics. While the system evolves infinitesimally under the system Hamiltonians, the bath undergoes a
finite propagation under fluctuations. This composite propagation results in a Markovian master equation with
a memory kernel in all dissipator terms \cite{chakrabarti2018b}. FRQME has been used as an essential tool
for theoretical analysis of drive-induced effects in quantum optics, quantum information processing, and
nuclear magnetic resonance (NMR) \cite{chatterjee_nonlinearity_2020, chanda2021, chanda2020, saha2022,
saha_effects_2022}.

For a driven dipolar dissipative system, in addition to the first-order effects, one can have dissipators from
a drive, dipolar relaxation terms, and their cross terms, all potentially much larger than the relaxation terms
from system-environment coupling \cite{chakrabarti_2022_creation}. Naturally, FRQME can describe the
dynamics of nearly closed systems at an intermediate timescale when the above condition is met. We shall
show that the dynamics of such a system is dominated by terms other than the relaxation terms, and the
journey to the final steady state may have multiple timescales.  

In the following, we analyze such a system as a prototype closed system by keeping the system-environment
coupling vanishingly small compared to the other terms in the equation. We obtain a set of quasi-conserved
quantities which aid in evolving the system from the initial state to the final steady state punctuated by one
or more prethermal plateau or an \emph{arrested state}. Each prethermal state is characterized by quasi-conserved
quantities and the decay of each plateau is associated with the breaking of these quasi-conservation laws. 

We consider two dipolar coupled spin-$1/2$ particles with the same Zeeman splitting. They are periodically
driven. The individual spins are weakly coupled with the external thermal environment that is undergoing
fluctuations with characteristic timescale $\tau_c$. Except for the fluctuations, the rest of the model
mimics the experimental condition at which Beatrez \etal observed a prethermal plateau
\citep{beatrez_floquet_2021}. 

The total Hamiltonian of the system and its environment is given by
\begin{eqnarray}\label{ham}
\mathcal{H}(t)&=& \hs^{\circ} + \hl^{\circ} +\hdr(t) + \hsl + \hdd + \hl(t).
\end{eqnarray}
Here, the first two terms are the free Hamiltonians of the system and local environment. $\hs^{\circ}$
represents the Zeeman interaction and the form is given by $\sum\limits_{i=1}^2\omega_\circ I_z^i$, where
$\omega_\circ$ is the Zeeman frequency. Here $I_{\alpha}^i = \sigma_{\alpha}^i/2, \,\alpha \in \{x,y,z\}$,
$\sigma_{\alpha}$ is $\alpha$ component of Pauli spin operators for spin-1/2 particles, and $i$ is the spin-index.
The analytical form of dipolar coupling in spherical tensor form is given by $\hdd = \sum\limits_{m=-2}^{2}
(-1)^m \omega_{d_{m}} \mathcal{T}^{m}_2$. Here, $\omega_{d_{m}}= \left(\mu_{\circ} \hbar \gamma^2/ 4 \pi
r^3\right) \mathcal{Y}^{-m}_2(\theta,\phi) $.  $\mathcal{Y}^{-m}_2(\theta,\phi)$ is the spherical harmonics
of rank $2$,  and $\mathcal{T}^{m}_2$ is the irreducible spherical tensor (of rank $2$ and order $m$).
$(\theta,\phi)$ is the polar and azimuthal angle of the average orientation of the dipolar vector
\textit{w.r.t.} the direction of the magnetic field. $\gamma$ is the gyro-magnetic ratio of the spin and
$\mu_{\circ}$ is the permeability constant. $\vec{r}$ is the average distance between the two spins. A
linearly polarized and weak periodic drive is applied in the transverse direction of the system. The
corresponding drive Hamiltonian is written as, $\hdr(t) = \sum\limits_{i=1}^2 \omega_1 I_{x}^i  \cos\omega t$.  

$\hsl$ depicts the system-environment coupling Hamiltonian, which provides the relaxation terms. $\hl(t)$
represents the spontaneous thermal fluctuation in the local environment which eventually provides a memory
kernel in the dissipator \cite{chakrabarti2018b}. The form is chosen as, $\hl(t) = \sum_i f_i(t)\vert \phi_i
\rangle \langle \phi_i \vert$, where, $\{\vert\phi_i\rangle\}$ are the eigenbasis of the environment.  $f_i$
are the independent Gaussian stochastic variables with zero mean and standard deviation $\kappa \,
\left(\kappa^2=\frac{1}{\tau_c}\right)$, here $\tau_c$ is the correlation time. This specific choice of
$\hl(t)$ ensures that the explicit presence of the thermal fluctuation doesn't destroy the equilibrium
density matrix of the environment, but it does destroy the coherences in the environment with a timescale
$\tau_c$. We assume that a clear timescale separation exists, i.e., $\tau_c$ is much smaller than the system's
timescale. Then one can construct a propagator that evolves the system infinitesimally and evolves the
environment under fluctuations by a finite amount. A coarse-graining followed by a cumulant expansion of the
fluctuation part of the propagator results in a Markovian quantum master equation with a memory kernel in
all its dissipators \cite{chakrabarti2018b}. For the case in hand,
the dynamical equation of the reduced system density matrix -- the FRQME -- in the interaction 
picture of the free Hamiltonians $\hs^{\circ} + \hl^{\circ}$, is given by,
\begin{eqnarray} \label{frqme}
\frac{d\rh}{dt}&=& -i \trl\Big[\heff(t),\rh \otimes\rl\Big]^{\rm sec}\nn\\
&&-\int\limits^{\infty}_0 d\tau \trl\Big[\heff(t),\Big[\heff(t-\tau),\rh \otimes\rl\Big]\Big]^{\rm
sec}e^{-\frac{\vert\tau\vert}{\tau_c}},
\label{eq:2}
\end{eqnarray}
where, $\heff(t)$ is the interaction representation of $\hsl +\hdr(t)+\hdd(t)$. $\rl$ is the
equilibrium density matrices of the environment. On the right-hand side, the `sec' denotes the secular
approximation \cite{cohen2004}. We note the presence of the exponential kernel
$\left(\exp(-t/\tau_c)\right)$, which gives a finite second-order effect of any local Hamiltonian (e.g.
$\hdr(t)$ and $\hdd $) along with $\hsl$. 

In the interaction picture, the drive and dipolar Hamiltonian can be written as the sum of the secular
(time-independent) and non-secular (time-dependent) parts. The analytical form of the secular parts is given
by, $\hdr^{\rm sec} = \omega_1 \sum\limits_i I_x^i$ and $\hdd^{\rm sec} =  \omega_{d_0}\mathcal{T}^{0}_2$. Similarly
the non-secular parts are written as, $\hdd^{\rm n} =\sum\limits_{m }  (-1)^m \omega_{d_{m}}
\mathcal{T}^{m}_2e^{-i m \omega_{\circ}t}\quad[\forall m\neq0]$, $\hdr^{\rm n}  = \sum\limits_{i}\omega_1 \left(
I_+^i e^{+2i\omega_{\circ}t} + I_-^i e^{-2i\omega_{\circ}t} \right)$.  In the superscript, $\rm "sec"$ denotes
the secular part, and $\rm "n"$ denotes the non-secular part of the Hamiltonian. We note that the effect of the secular
components is much stronger than the non-secular component in the dynamics, as the secular parts are present
in the first order commutator, whereas, the non-secular components are only present in the dissipators.

In Liouville space, the dynamical equation can be written as $\frac{d \hat{\rh}(t)}{dt}=
\hat{\mathcal{L}}\hat{\rh} (t)$. For our case, $\hat{\mathcal{L}}$ is a $16 \times 16$ matrix and
$\hat{\rh}$ is a $16 \times 1$ column matrix. The detailed form of $\hat{\mathcal{L}}\hat{\rh}$ is given by,
$\hat{\mathcal{L}}\hat{\rh} = \left(\hat{\mathcal{L}}_{sec}  + \hat{\mathcal{L}}_{n}  +
\hat{\mathcal{L}}_{\text{\tiny{SL}}}\right)\hat{\rh}$. The three terms, $\hat{\mathcal{L}}_{sec}$,
$\hat{\mathcal{L}}_{n}$,  $\hat{\mathcal{L}}_{\text{\tiny{SL}}}$ denote the Liouvillian corresponding to the
secular parts, the non-secular parts, and system-bath coupling Hamiltonian respectively.  

Instead of working with a full density matrix, we chose to rewrite the equation in terms of the expectation
values of the relevant spin observables, which can be easily interpreted to gain insight into the dynamics.
Normally, a two-spin density matrix can be written by using fifteen observables. For a symmetric case, the 
number of observables is reduced to nine. The representation of $\rh(t)$ is given by,
\begin{eqnarray}\label{obs}
\rh (t) &=& \sum\limits_{\alpha, \beta }  A_{\alpha \beta} I_{\alpha} \otimes I_{\beta},
\end{eqnarray}
where, $\alpha$, $\beta$ can take values from $\{x,y,z,d\}$, and $I_d = 2\times 2$ identity matrix. 
Here, we further define $M_i = A_{id}  + A_{di}$, $M_{ii} = A_{ii}$, $M_{ij} = A_{ij} + A_{ji}$ $\forall i\neq j$ and $i,j \neq d$.

\begin{figure}[htb]
\includegraphics[width=0.48\linewidth]{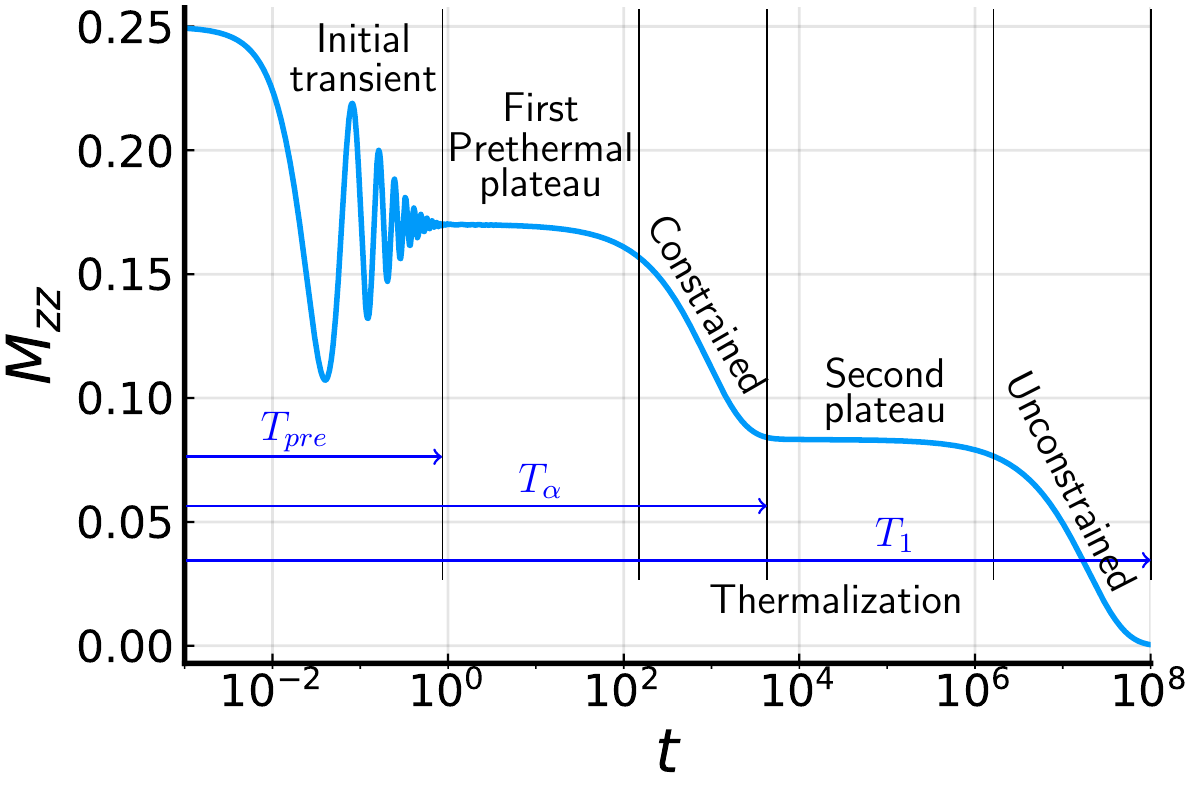} 
\caption{The plot shows the complete dynamics of the observable $M_{zz}$ as a function of $t$ in log-scale.
The plot was generated by solving FRQME in Eq. (\ref{frqme}) for  $\omega_1 = 2\pi \times 5$ KHz, $\vert 
\omega_{d_m}\vert = 2\pi \times 5 $ KHz, $\omega_{\circ} = 2\pi \times 10 $ MHz, $\tau_c = 1$ $\mu$s, 
$\omega_{\text{\tiny{SL}}} = 2\pi \times 3\times 10^{-2}$ KHz, and $M_{zz}\stoo = \frac14$. The choice of
parameters ensures that regular relaxation comes into play long after the other dissipators have completed
their part. The dynamics clearly show the emergence of multiple prethermal states and the cascaded journey
to the final steady state. $T_{pre}$ shows the timescale at which the initial transients die out and the
first prethermal state emerges. This state further decays under a constrained thermalization process, and the
second prethermal plateau emerges after $T_{\alpha}$. The regular relaxation with a longer timescale $T_1$
finally takes the system to a steady state.}
\label{fig-1}
\end{figure}

First, we show the complete solution of Eq. \ref{eq:2} in terms of one of the observables defined in Eq.
\ref{obs} for the initial condition, $\rh \stoo = \vert 11 \rangle \langle 11 \vert$ (i.e. $M_{zz}\stoo =
\frac14$) in fig. \ref{fig-1}. The figure caption contains the parameters used for generating the plot. We
observe a cascaded journey of this observable from the initial value to the final steady state. We shall
analyze the cascaded evolution by systematically including various interactions with decreasing strengths.

We note that Chakrabarti \etal has recently described the emergence of prethermal plateau
\cite{chakrabarti_2022_creation}, and hence we provide a brief review here. One needs to consider the
strongest Liouvillian ($\mathcal{L}_{sec}$) to understand the dynamics.  Only the effect of the secular
parts are considered here $\left(\mathcal{H}_{tot}^{sec} = \hdr^{sec} + \hdd^{sec}\right)$ . The explicit
form of $\hat{\mathcal{L}}_{sec}$ is given below,  
\begin{eqnarray}\label{dynamics-I}
\mathcal{L}_{sec}\rh(t)&=& -i\left[\mathcal{H}_{tot}^{sec}, \rh(t)\right] -  \tau_c\left[ \mathcal{H}_{tot}^{sec},\left[\mathcal{H}_{tot}^{sec},\rh(t) \right] \right]
\end{eqnarray}
Drive and dipolar interaction can appear in both first order and second order of the above Eq.
(\ref{dynamics-I}). Cross relaxation occurs due to the cross terms of $\hs(t)$ and $\hdd$ in the second
order.
 
The dynamical equation for the collective coherence ($M_x$) and corresponding coupled observables which are
connected to the quasi-conserved quantities are written as,
\begin{eqnarray}
\dot{M}_{x} &=& -\frac{9}{4} \omega_{d_0}^2 \tau_c  M_{x} + 6 \omega_1 \omega_{d_0} \tau_c M_{zz} -6 \omega_1 \omega_{d_0} \tau_c M_{yy} -3\omega_{d_0} M_{yz} \label{eq-pre1} \\
\dot{M}_{zz} &=&  \frac{3}{4}\omega_1 \omega_{d_0} \tau_c M_{x} -2\omega_1^2 \tau_c M_{zz} +  2\omega_1^2 \tau_c M_{yy} + \omega_1 M_{yz}\label{eq-pre2} \\
\dot{M}_{yy} &=&  -\frac{3}{4}\omega_1 \omega_{d_0} \tau_c M_{x} +2\omega_1^2 \tau_c M_{zz} -  2\omega_1^2 \tau_c M_{yy} - \omega_1 M_{yz}\label{eq-pre3} \\
\dot{M}_{yz} &=& \frac{3}{4}\omega_{d_0} M_{x} -2 \omega_1 M_{zz} +2\omega_1 M_{yy} -(4 \omega_1^2 + \frac{9}{4}\omega_{d_0}^2 \tau_c)M_{xy} \label{eq-pre4}
\end{eqnarray}

We note that the above equations show several constants of motion. For example, from the above, we infer,
$M_{yy}+M_{zz}$, $3\omega_{d_0}M_{zz}+\omega_1  M_{x}$ are constants. From the other equations (not shown),
we also have $\dot{M}_{xx} = 0$, and hence $M_{xx}$ is also a conserved quantity. However, as one includes
other non-secular terms, the quantities may not be conserved. But being weaker than secular parts, the
effect of non-secular terms becomes evident in the later part of the dynamics. As such, these quantities
appear to be constant for a while and then undergo changes. Following Peng \emph{et al.} we term these
quantities as quasi-conserved \cite{peng_floquet_2021}. Since the presence of the coherence $M_x$ as an
arrested state was observed recently by Beatrez \emph{et al.} \cite{beatrez_floquet_2021}, we name the
quasi-conserved quantity $3\omega_{d_0}M_{zz}+\omega_1 M_{x}$ as the prethermal order.  Similarly, we have
another quasi-conserved quantity $M_{xx}+M_{yy}+M_{zz}$; this is the expectation value of $J^2 \propto
\vec{\sigma}^1.\vec{\sigma}^2$, and we name it the dipolar order. The quasi-conserved quantities can also be
analyzed by using the eigen-spectrum of $\mathcal{L}_{sec}$. It has four zero-eigenvalues. One of the zero
eigenvalues corresponds to the preservation of the trace. The three other zero eigenvalues correspond to the
above three quasi-conserved quantities. 

While solving the equation, the initial condition is chosen as $M_{zz}\stoo = M_{\circ}$ and other observables
as zero. The solution of $M_{zz}(t)$ from Eq.  (\ref{dynamics-I}) is given by,
\begin{eqnarray}
M_{zz}^{\text{\tiny{pre}}}(t) &=&  \frac{M_{\circ}}{4}\left(1 - \frac{2\omega_1^2}{\kappa_1^2}\left(1 -
\cos(\kappa_1 t)e^{-t \kappa_1^2 \tau_c}\right)\right)
\end{eqnarray}
where, $\kappa_1^2 = 4\omega_1^2 + 9\omega_{d_0}^2/4$.
Therefore, $M_{zz}\sts  = \frac{M_{\circ}}{4}(1-2\omega_1^2/\kappa_1^2)$ . The above solution
shows that the initial transient phase oscillates with a frequency $\kappa_1 $ and it reaches the
prethermal steady state $(M_{zz}\sts)$ in a characteristic time-scale $T_{pre} = 1/\left(\tau_c \kappa_1^2
\right)$. In Fig. \ref{fig-1}, $T_{pre}$ indicates the timescale of the system to reach the prethermal state
from the initial transient phase. 

Next, we include $\hat{\mathcal{L}}_{n}\hat{\rh}$ in the dynamics, whose form is given by,
\begin{eqnarray}\label{eq-3}
\mathcal{L}_{n}\rh= -i[\mathcal{H}_{bs} + \mathcal{H}_{ds}, \rh ] + \mathcal{D}_1 \rh + \mathcal{D}_2 \rh 
\end{eqnarray}
The shift terms are the Kramers-Kronig pairs of the second-order dissipative terms of the non-secular parts.
They result in the re-normalization of the Zeeman frequency. $\mathcal{H}_{bs}$ is called the
Bloch-Siegert shift, $\mathcal{H}_{bs}= 2\omega_1^2 \omega_{\circ} \tau_c \mathcal{Z}(2)  I_z^i $, similarly
$\mathcal{H}_{ds}$ represents the second-order dipolar shift; the expression is given by $\mathcal{H}_{ds} =
(\omega_{d_1}^2 \omega_{\circ} \tau_c \mathcal{Z}(1) + 2\omega_{d_2}^2 \omega_{\circ} \tau_c
\mathcal{Z}(2))I_z^i$. Here $\mathcal{Z}(m) = \tau_c/\left(1+(m \omega_{\circ}\tau_c)^2 \right)$. These
shift terms do not contribute to the decay rate and minimally affect the steady-state configuration, so we
ignore them in the rest of the analysis. The explicit forms of the dissipative operators in Eq. (\ref{eq-3})
can be written as,
\begin{eqnarray}
\mathcal{D}_1 \rh  &=& \sum\limits_{m=-2}^2 \vert\omega_{d_m}\vert^2 \mathcal{Z}(m) \left[2 \mathcal{T}_2^{-m} \rh \mathcal{T}_2^{m} - \left\{\mathcal{T}_2^{m}\mathcal{T}_2^{-m}, \rh \right\}\right]\quad [\forall m \neq 0]\nn\\
\mathcal{D}_2 \rh  &=& \sum\limits_{i,j=1}^2 \omega_1^2\mathcal{Z}(2)  \left[2 I_{\pm}^i \rh I_{\mp}^j-\left\{ I_{\mp}^j I_{\pm}^i, \rh \right\} \right] 
\end{eqnarray}

We analyze the above Eq. (\ref{eq-3}) using the observables defined earlier in Eq. (\ref{obs}). The
super-operator, $\hat{\mathcal{L}_{sec}}+\hat{\mathcal{L}_n}$ has two zero eigenvalues. 
In the presence of the non-secular terms, two of the
earlier four quasi-conserved quantities will no longer be conserved. The prethermal state does not survive in this
regime, as $3\omega_{d_0} \dot{M}_{zz} +  \omega_1  \dot{M}_{x}  \neq 0$. Also, we have $\dot{M}_{xx} \neq
0$ and, $\dot{M}_{zz} + \dot{M}_{yy} \neq 0$.  But the dipolar order still survives, ($\dot{M}_{xx}+
\dot{M}_{yy}+\dot{M}_{zz} =0$). Therefore this regime shows a constrained thermalization, as the initial
memory partly survives. The resulting equation of the observables are,
\begin{eqnarray}\label{eq-4}
\dot{M}_x &=& - (2q_1 + \frac{5}{2}p_1 + p_2) M_x + \text{terms in Eq. (\ref{eq-pre1})} \nn\\
\dot{M}_{zz} &=& -(8 q_1 + 2 p_1)M_{zz} +(4q_1 +p_1)(M_{xx} + M_{yy})+ \text{terms in Eq. (\ref{eq-pre2})}\nn\\
\dot{M}_{yy}&=& -(4q_1 + p_2 + p_1)M_{yy} + p_2 M_{xx} + (4q_1 + p_1)M_{zz}+\text{terms in Eq. (\ref{eq-pre3})}\nn\\
\dot{M}_{xx}&=& -(4q_1 + p_2 + p_1)M_{xx} + p_2 M_{yy} + (4q_1 + p_1)M_{zz}
\end{eqnarray}
Here $q_1 = \omega_1^2\mathcal{Z}(2) $ and $p_m = \vert\omega_{d_m}\vert^2 \mathcal{Z}(m) $. In the limit of 
$\omega_{\circ} \tau_c >1$, the solution of transverse magnetization mode $M_{zz}(t)$ is written as, 
\begin{eqnarray}\label{mx-2}
M_{zz}(t) &\approx& \frac{M_{\circ}}{3}  - e^{-t/T_{\alpha}}\left(\frac{M_{\circ}}{3} -  M_{zz}^{\text{\tiny{pre}}}(t)\right) \\
\frac{1}{T_{\alpha}} &=&3\left(p_1 + 4q_1 \right)\nn
\end{eqnarray}
Hence, the above solution of $M_{zz}(t)$ gives the notion of decay of prethermal order. 
$T_{\alpha}$ is the decay rate due to the non-secular part. 

\begin{figure*}[htb]
 \raisebox{3cm}{\normalsize{\textbf{(a)}}}\hspace*{-1mm}
\includegraphics[width=0.36\linewidth]{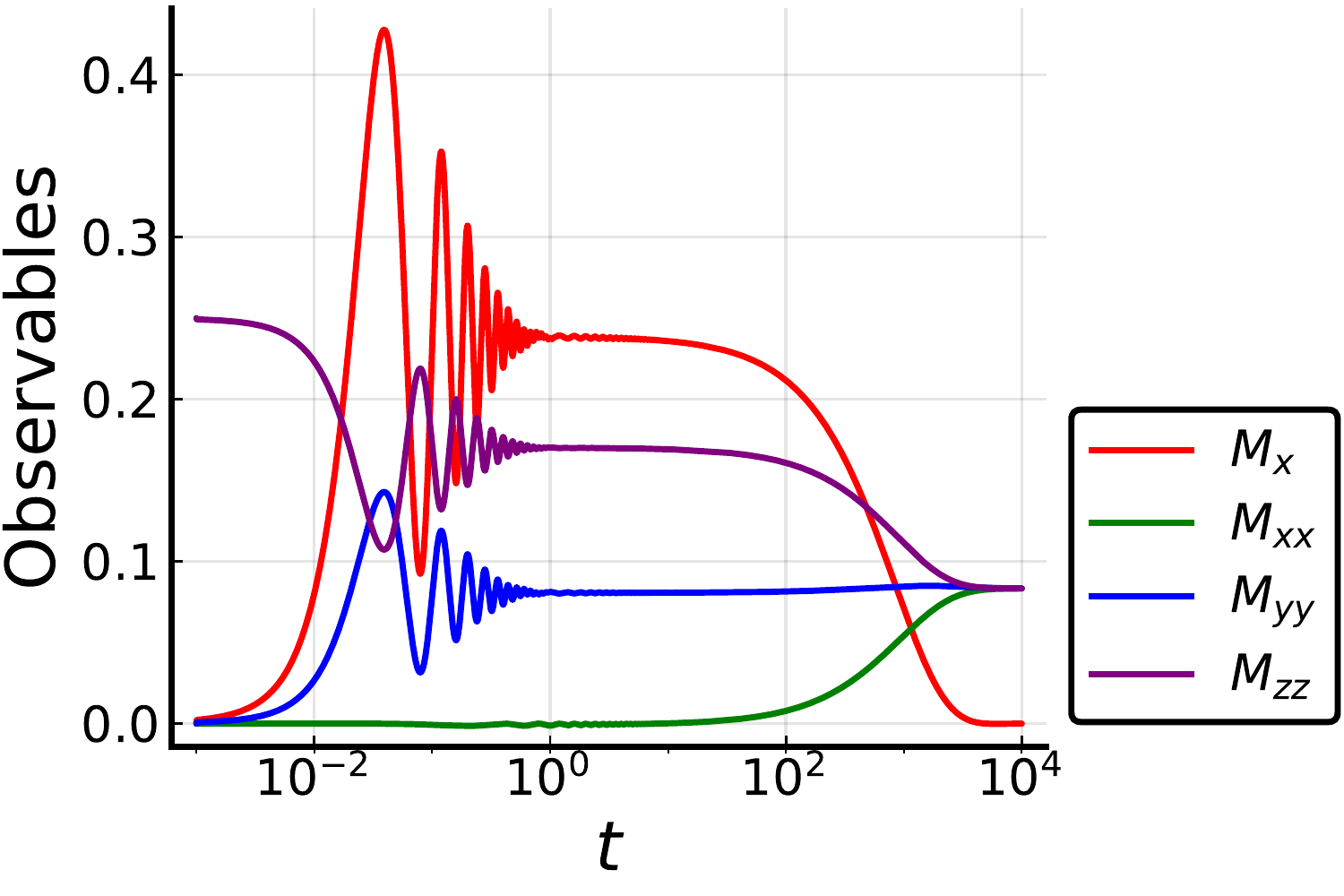} 
\hspace*{3mm}
\raisebox{3cm}{\normalsize{\textbf{(b)}}}\hspace*{-1mm}
\includegraphics[width=0.36\linewidth]{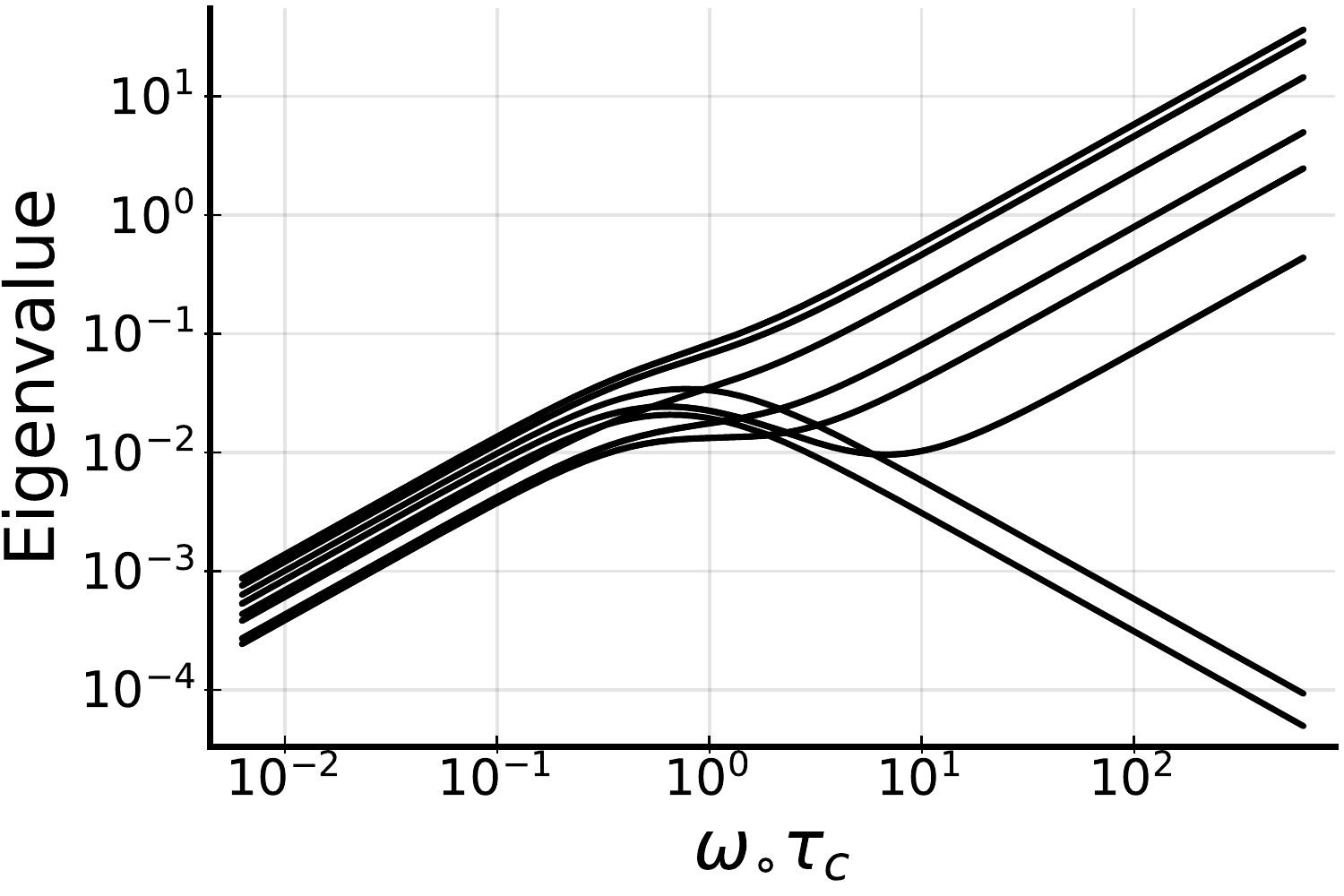} 
 \raisebox{3cm}{\normalsize{\textbf{(c)}}}\hspace*{-1mm}
\includegraphics[width=0.36\linewidth]{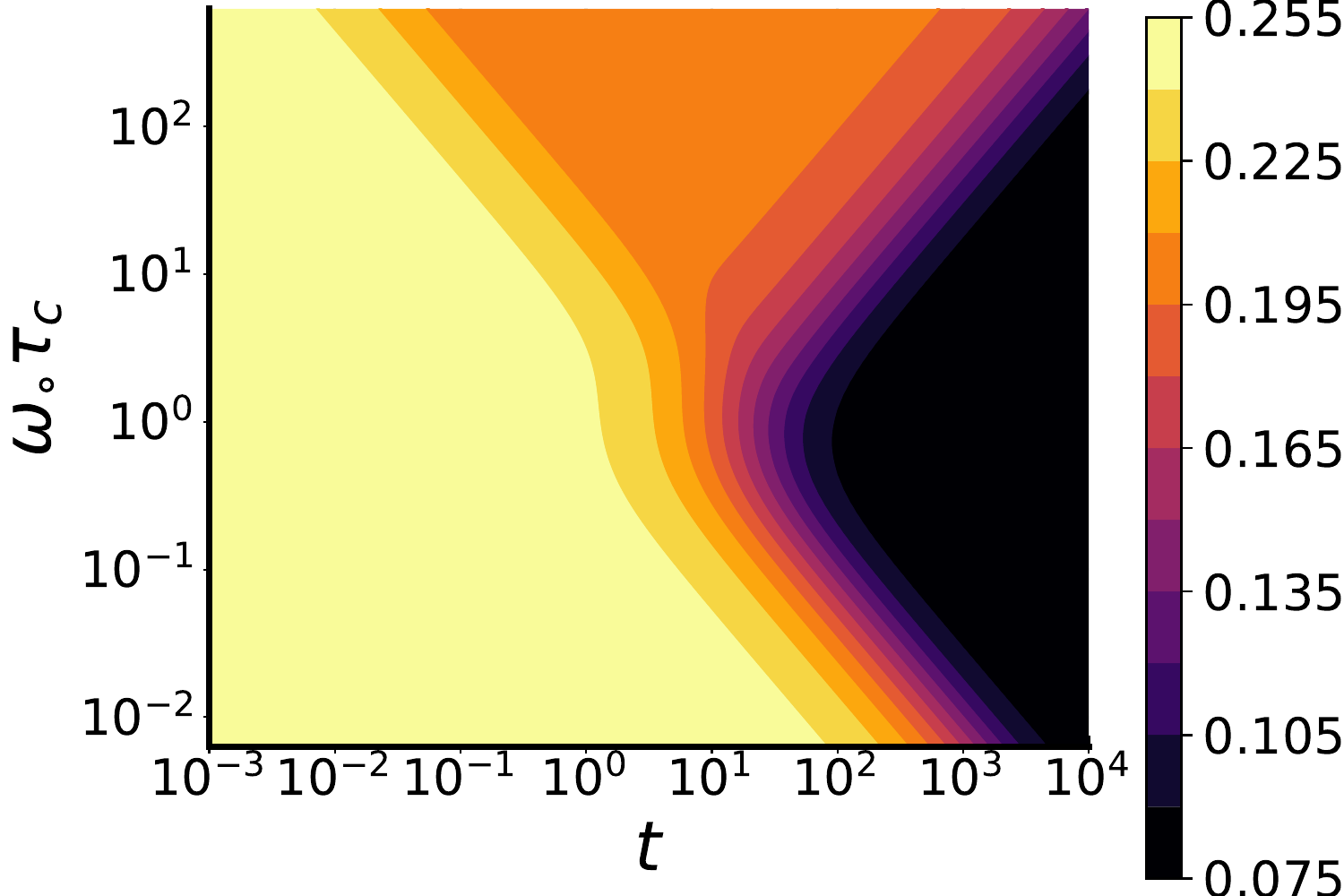} 
\raisebox{3cm}{\normalsize{\textbf{(d)}}}\hspace*{-1mm}
\hspace*{3mm}
\includegraphics[width=0.38\linewidth]{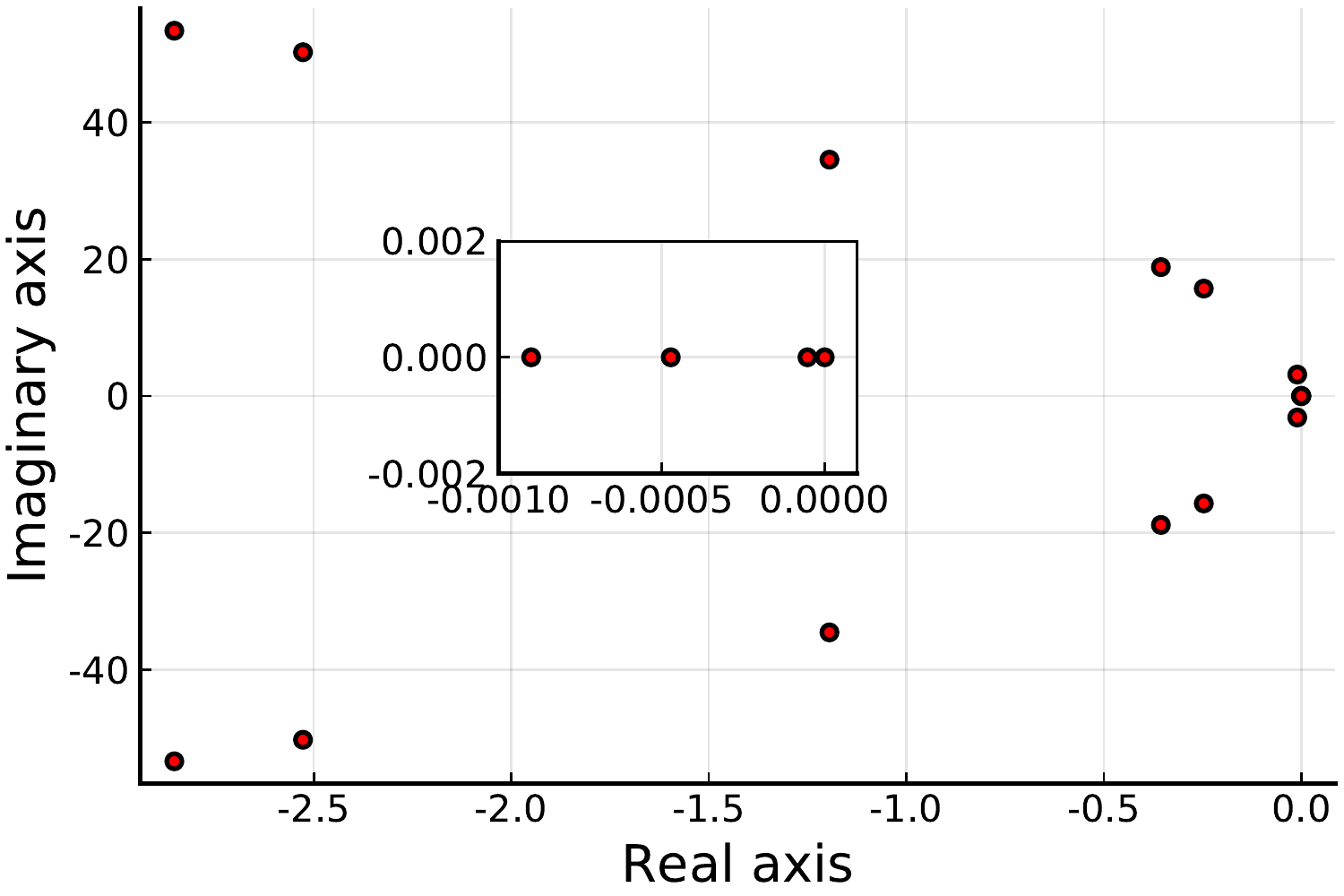} 
\caption{Subfigure (a) shows the plot of $\{M_x,\, M_{xx},\, M_{yy},\, M_{zz}\}$ vs $t$. All the plots are
done by numerically solving Eq. (\ref{dynamics-I}) and  Eq. (\ref{eq-3}) simultaneously. The values of the
parameters are given below, $\omega_1 = 2\pi \times 5$ KHz, $\vert \omega_{d_m}\vert = 2\pi \times 5 $ kHz,
$\omega_{\circ} = 2\pi \times 10 $ MHz, $\tau_c = 1$ $\mu$s. It shows that in the intermediate timescale,
the prethermal order is preserved, but after a characteristic time, it decays. First two eigenvalues of total
$\hat{\mathcal{L}}$ (excluding $\hsl$) are zero. Modulus of the other $14$ eigenvalues are plotted as a
function of $\omega_{\circ}\tau_c$ in subfigure (b). For $\omega_{\circ}\tau_c>1$, The separate existence of
the lowest two eigenstates from others is clearly visible. Whereas, for $\omega_{\circ}\tau_c\leqslant 1$,
they merge.  Subfigure (c) shows the filled contour of $M_{zz}(t)$ as a function of $\omega_{\circ}\tau_c$
and $t$.  $\omega_{\circ} \tau_c \approx 1$ is the critical point of the dynamics. For $\omega_{\circ}
\tau_c > 1$, the orange (color online) triangular region shows the prethermal plateau. For $\omega_{\circ}
\tau_c < 1$, no such plateau forms. For (b) and (c), the effect of the imaginary part of the eigenvalues is
neglected. Subfigure (d) shows the numerical plots for the distribution of the sixteen eigenvalues of
$\hat{\mathcal{L}}_{sec}  + \hat{\mathcal{L}}_{n}  + \hat{\mathcal{L}}_{\text{\tiny{SL}}}$ in the real and
imaginary axis. Here $P_- = 1.6 \times 10^{-5}$, $P_+ = 0.4 \times 10^{-5}$ The lowest four eigenvalues are
close to zero and are merged. The inset in the middle of the plot shows those four eigenvalues with an enlarged
scale. The first two eigenvalues in the inset from the left are responsible for the prethermal plateau. The
third one (from the left) shows the notion of constrained thermalization. The right one leads the system to
the final steady state. }
\label{fig-4}
\end{figure*}

The dynamics of the other observables connected to prethermal order and dipolar order $\{M_x,\,M_{xx},\,
M_{yy},\, M_{zz}\}$ are also shown in Fig.\ref{fig-4}(a). The observables $\{M_{yy},\, M_{zz}\}$ get non-zero steady-state values in the prethermal time-regime, and $M_{xx}$ remains zero. Then at a later time,
they begin to evolve. The constrained thermalization is shown in the time regime $T_{\alpha}$ of
Fig.{\ref{fig-1}}.  The aforementioned terms in Eq. (\ref{eq-3}) give an effective decay in the dynamics. The
decay time is denoted as $T_{\alpha}$. The expression of the decay rate is given in Eq. (\ref{mx-2}). It is
generally expressed as $1/T_{\alpha} \propto  \frac{1/T_{pre }}{1 + (\omega_{\circ}\tau_c)^2}$. If one
begins with $M_x\stoo \neq 0$ and other observables are zero, then it replicates the spin-locking
phenomena in nuclear magnetic resonance (NMR) and the recent experiment by Beatrez \etal
\cite{beatrez_floquet_2021}. 

For a very weak system-bath interaction (long $T_1$ process), the timescale corresponding to the decay of quasi-equilibrium
state is denoted as $T_{1 \rho}$ in the spin-temperature theory of NMR \cite{slichter1996}. In our case,
$T_{\alpha}$ is equivalent to $T_{1 \rho}$. The necessary condition for having a prethermal
plateau is $T_{\alpha} \gg T_{ pre}$. Therefore, to fulfill this condition, we must have $\omega_{\circ}
\tau_c \gg 1$. In this limit, $1/T_{pre} \propto \tau_c$ and $1/T_{\alpha} \propto 1/\tau_c$.

It is known that, in the case of the dynamics of a dissipative system, the eigenvalues of
$\mathcal{\hat{L}}$ must lie in the non-positive plane. The imaginary part of the eigenvalue comes from
the unitary process due to the first-order terms and shift terms. On the other hand, the real part depicts
the non-unitary dissipation \cite{albert_symmetries_2014}. The lifetime of the prethermal plateau is
independent of the unitary process, so we neglect the imaginary part of the eigenvalue of $\mathcal{\hat{L}}$
for the rest of the analysis. 

In this regime, $\mathcal{L}$ has two zero eigenvalues. Others are the decaying modes. The decay times are
obtained by the inverse of the modulus of the eigenvalues. In Fig. \ref{fig-4}(b), we have plotted the modulus
of the eigenvalues as a function of the dimensionless quantity $\omega_{\circ} \tau_c$. It shows that for
$\omega_{\circ} \tau_c>1$, two eigenvalues are much smaller than the others. When the remaining twelve modes decay,
those two modes still remain alive, and the system has a prethermal plateau of a finite lifetime. Whereas, for
$\omega_{\circ} \tau_c<1$, all the fourteen modes decay nearly at the same time. Hence, the plateau will not
survive in this regime.

The solution of $M_{zz}(t)$ as a function of time and for the specific choice of $\omega_{\circ} \tau_c$ is
shown in the contour plot (Fig. \ref{fig-4}(c)). It depicts clearly a critical point at $\omega_{\circ}
\tau_c = 1$. For $\omega_{\circ} \tau_c>1$, there exist two distinct decay rates. The prethermal plateau is
clearly visible in the regime. As we increase the value of $\omega_{\circ} \tau_c$, the length of the
plateau increases as the differences between the eigenvalues grow. For, $\omega_{\circ} \tau_c<1$, the
plateau ceases to exist. This criticality is reminiscent of the behavior of the relaxation rates due to
spin-lattice relaxation as a function $\omega_{\circ} \tau_c$, as reported by Bloembergen and others
\cite{bloembergen_relaxation_1948, bloembergen_nuclear_1947}. They found that for high motional narrowing,
$\omega_{\circ} \tau_c < 1$, and for the other regime (large macromolecules or solids) $\omega_{\circ}
\tau_c > 1 $.

For the chosen parameters, the effect of $\hsl$ is prominent after the system reaches the unconstrained 
thermalization regime. We are not using any specific form of system-bath coupling; we introduce the 
relaxation rates in terms of known parametric form in the equation, as shown below.
The form of $\hat{\mathcal{L}}_{\text{\tiny{SL}}}$ is given by,
\begin{eqnarray}
\mathcal{L}_{\text{\tiny{SL}}}\rh  &=& -i\left[\mathcal{H}_{\text{\tiny{lamb}}},\rh\right] + \mathcal{D}_3 \rh  
\end{eqnarray}
$\mathcal{H}_{\text{\tiny{lamb}}}$ is known as Lamb shift. It also has a minimal contribution to the steady state. 
The form is $D_3 \rh$ is given by, 
\begin{eqnarray}
D_3 \rh &=& \sum\limits_{i=1}^2P_{\mp}\left[  2I_{\mp}^i \rh I_{\pm}^i - \{I_{\pm}^i I_{\mp}^i, \rh\}
\right],
\end{eqnarray}
where $P_-, P_+$ is defined as the probability of the downward and upward transition due to $\hsl$
\cite{breuer2002}. The form of $P_-$, $P_+$, and the dynamical equation in terms of the observables are
shown in our earlier work \cite{saha_effects_2022}. In the presence of only $\hsl$ (excluding $\hs(t)$,
$\hdd$), the final density matrix is diagonal. For this case, the steady-state solution is given by,
$M_z\sts = \frac{P_--P_+}{P_-+P_+}$, $M_{zz}\sts = \left(M_z\sts\right)^2/4$, and $T_{1} = 1/(P_+ +P_-)$.

The total Liouvillian ( $\hat{\mathcal{L}}$) has a single zero eigenvalue, which means the steady
state is unique and it has no initial memory. Under the evolution of $\hsl$, the dipolar order is broken,
$\left[\dot{M}_{xx} + \dot{M}_{yy} + \dot{M}_{zz} \neq 0\right]$. Therefore the system reaches a final
steady state under the evolution of full $\mathcal{L}$. The distribution of the eigenvalues is plotted in
fig. \ref{fig-4}(d). In the inset, the first three eigenvalues (from the left in the inset) decay at a
longer time scale. Among them, the first two are responsible for the prethermal plateau, and the third one is
for constrained thermalization, as only the dipolar mode survives after the prethermal state decays. The
zero eigenvalue stands for the steady state; it appears when all fifteen eigenmodes decay.

The decay rate of the dipolar order is much slower than the prethermal order. This kind of
quasi-conservation law was recently predicted by Peng \etal in kicked dipolar model \cite{peng_floquet_2021}.
The characteristic decay time is denoted as $T_1$ [$1/T_1 \propto \frac{\omega_{sl}^2 \tau_c}{1 +
(\omega_{\circ}\tau_c)^2}$]. For $T_{\alpha} \gg T_1$, there exists another plateau in the system, which is
shown in Fig. \ref{fig-1}. For $\omega_{\circ} \tau_c < 1 $, [$T_{\alpha} \approx T_{pre}$], the initial
transient phase skips the prethermal plateau and directly goes to the second plateau, whereas, for $T_1 \gg
T_{\alpha}, T_{pre}$, the system directly arrives at the steady state, as there is no existence of
prethermalization. 

\begin{table}[htb]
\begin{center}
\caption{comparison of different dynamical phases}
\begin{tabular}{ p{5.0cm} |p{2.5cm} |p{4.3cm}|  p{3.50cm}   }
 Dynamical Phase  & Liouvillian & Quasi-conserved quantity & Dark state\\
 \hline
 \hline
  Prethermalization  &  $\hat{\mathcal{L}}_{sec} $   &(a) $3\omega_{d_0} M_{zz} +  \omega_1  M_{x}$, & All eigenstates   \\
 & &(b) $M_{xx}$, (c) $M_{yy}+M_{zz} $ & corresponding to (a) \\
  Constrained thermalization & $\hat{\mathcal{L}}_{sec}  + \hat{\mathcal{L}}_{n} $ & (a) $M_{xx}+M_{yy}+M_{zz} $ & Only Singlet state  \\
  Unconstrained thermalization & $\hat{\mathcal{L}}_{sec}  + \hat{\mathcal{L}}_{n}  + \hat{\mathcal{L}}_{\text{\tiny{SL}}}$ &  None & None  \\
 \end{tabular}
\label{table-1}
\end{center}
\end{table} 

The cascaded dynamics can be analyzed by the existence of the dark states in each regime. The
non-equilibrium steady state ($\rho_{\infty}$) is defined as the null vector of the Liouvillian
($\mathcal{L}\rho_{\infty}= 0 $). If $\rho_{\infty}$ is a pure density matrix ($\rho_{\infty}=\vert
\psi_\infty \rangle \langle \psi_\infty \vert$), then $\vert \psi_\infty \rangle$ is defined as dark state
\cite{buca_note_2012}. Here, in the prethermal regime, all the eigenstates of the observables corresponding
to the prethermal order act as a dark state. When the prethermal plateau decays, three of the previous four
states are no longer dark states. In the constrained thermalization regime, only the Bell singlet state
($\frac{1}{\sqrt{2}}(\vert 10 \rangle  - \vert 01 \rangle)$) is acting as a dark state. Finally, when the
system reaches unconstrained thermalization, there exists no dark state in the system. Table \ref{table-1}
shows the summary of the above description.

A long prethermal plateau is useful for designing sensitive magnetometers and quantum-sensors
\cite{beatrez_floquet_2021}. The necessary condition for this is $T_{pre} \ll T_{\alpha} \ll T_1$ or $\omega_{\circ}
\tau_c \gg 1$. One can design a system to have a strong dipolar coupling and can apply a strong drive to
achieve this condition at any given temperature. We show here the simple case of a
dipolar-coupled two-spin system, whereas in the actual experiment, a dipolar network is considered. 
Our results show good agreement with the recent experiments by Beatrez \etal, and it can be extended to the
dipolar network.   
 
In summary, we present here the complete dynamical theory of periodically driven dissipative dipolar system
using FRQME. We show that the journey to the equilibrium for such systems is a multi-stage process. The
initial transient phase is governed by the first-order unitary processes. The second-order cross-relaxation
process leads to the prethermal state. This prethermal state further decays due to the non-secular part of
the drive and dipolar Hamiltonian. We calculate an effective decay rate that predicts the lifetime of such a
state. We show that there exists a critical limit to the temporal correlation of the local environmental
fluctuation. Beyond the limit, the prethermal plateau does not survive. We also observe that there exists
another quasi-conserved quantity, even after the prethermal state vanishes. Finally, we discuss the effect
of the system-bath coupling, which leads the system to the final steady state. The analysis in the above has
been carried out using specific values of system parameters to demonstrate the possible existence of
multiple prethermal plateaus. However, we note that for strong system-environment coupling, there may not be
a well-defined thermal plateau as the timescales of various may overlap with each other. We envisage that
this theoretical exposition of complete dynamics of a driven dissipative dipolar system will pave the way to
a better understanding of the thermalization problem.

The authors thank Prasanta K. Panigrahi, Arpan Chatterjee, and Arnab Chakrabarti for their insightful
comments and helpful suggestions. SS gratefully acknowledges University Grants Commission for a research
fellowship (Student ID: MAY2018- 528071).

\bibliographystyle{apsrev4-1}
\bibliography{ref2}

\end{document}